\documentclass[pre,aps,showpacs,twocolumn]{revtex4}
\usepackage{graphicx}
\usepackage{dcolumn}
\usepackage{bm}
\usepackage{amsmath,amssymb}
\usepackage{color,soul}

\begin{document}

\title{Enhanced magnetoelectric effect of exactly solved spin-electron model on a doubly decorated square lattice in vicinity of a continuous phase transition}
\author{Hana \v Cen\v carikov\'a$^{1}$ and Jozef Stre\v{c}ka$^2$}
\affiliation{
$^1$Institute of Experimental Physics, Slovak Academy of Sciences, Watsonova 47, 040 01 Ko\v {s}ice, Slovakia \\ 
$^2$ Department of Theoretical Physics and Astrophysics, Faculty of Science, P. J. \v{S}af\'{a}rik University, Park 
Angelinum 9, 040 01 Ko\v{s}ice, Slovakia}

\begin{abstract}
Magnetoelectric properties of a coupled spin-electron model on a doubly decorated square lattice in an external electric field applied along the crystallographic axis \textbf{[11]} are rigorously examined with the help of generalized decoration-iteration transformation. The phase diagram, spontaneous magnetization and electric polarization are exactly calculated  and their dependencies are comprehensively investigated under a concurrent influence of temperature and electric field. It is found that the electric field mostly stabilizes at zero temperature the spontaneous antiferromagnetic order with respect to the ferromagnetic one. At finite temperatures the external electric field gradually suppresses a spontaneous ferromagnetic (antiferromagnetic) order emergent close to a quarter (half) filling. An enhanced magnetoelectric response is detectable in vicinity of a continuous phase transition at which the spontaneous magnetization vanishes and the electric polarization displays a weak-type singularity. It is demonstrated that reentrant phase transitions of the ferromagnetic or antiferromagnetic phase may be induced at moderate values of the electric field, which simultaneously produces a sharp kink in a critical line of the ferromagnetic phase nearby a quarter filling. 
\end{abstract}


\maketitle
\section{Introduction}\label{sec:introduction}
Magnetoelectric effect currently attracts a great deal of attention though the early discovery of this cooperative phenomenon dates back to pioneering work by Curie more than a century ago \cite{cur94}. The main reason for this revival of interest lies in an immense application potential hidden in the magnetoelectric effect \cite{fie05,tok14}. The term magnetoelectric effect signifies a dependence of the magnetization on an electric field and a dependence of the electric polarization on a magnetic field. Sizable technological applications of magnetoelectric effect in the form of switches or data-storage devices are quite obvious, but they are unfortunately limited by a relatively small response of the magnetization (polarization) with respect to the external electric (magnetic) field (see Refs. \cite{fie05,tok14} and references cited therein).  Although the phenomenological theory of magnetoelectric effect has been developed by Dzyaloshinskii more than a half century ago \cite{dzy60} an incomplete understanding of the magnetoelectric effect at a microscopic level precludes an efficient enhancement of the magnetoelectric response. Exactly solved lattice-statistical models, which would provide a deeper understanding of the magnetoelectric effect and all its consequences, are therefore highly desirable. 

To the best of our knowledge, there are only a few rigorous studies of the magnetoelectric effect for one-dimensional Heisenberg \cite{bro13} and XX \cite{men15,bar18} spin chains, as well as, zero-dimensional Hubbard pair \cite{bal17,bal18,bas18} and cubic \cite{sza18} clusters. However, these rigorous studies cannot bring insight, because of their low-dimensionality, into the magnetoelectric response in a close vicinity of temperature-driven phase transitions. The main goal of the present work is therefore to fill in this gap when considering a coupled spin-electron model on a doubly decorated square lattice, which exhibits a nontrivial criticality at finite temperatures notwithstanding of it exact solvability \cite{tan10,str09,gal11,dor14,str15,cen16,cen18}. To achieve this goal, we will extend a coupled spin-electron model on a doubly decorated square lattice introduced in Refs. \cite{dor14,str15} by considering the external electric field applied along the crystallographic axis \textbf{[11]} and rigorously calculate the spontaneous magnetization as well as the electric polarization under a concurrent influence of the external electric field and temperature. It will be demonstrated hereafter that the investigated spin-electron model indeed exhibits an enhanced magnetoelectric effect in a close vicinity of a continuous phase transition. 

The paper is organized as follows. The coupled spin-electron model on a doubly decorated square lattice in presence of the external electric field will be introduced in Sec.~\ref{sec:model} together with a few basic steps of its exact treatment. The most interesting results for the ground-state and finite-temperature phase diagrams will be discussed along with typical temperature and electric-field variations of spontaneous magnetization and electric polarization in Sec.~\ref{sec:result}. The paper ends up with a brief summary of the most important findings and future outlooks presented in Sec.~\ref{sec:conc}.

\section{Model and Method}
\label{sec:model}

\begin{figure}[h!]%
{\includegraphics[width=.35\textwidth]{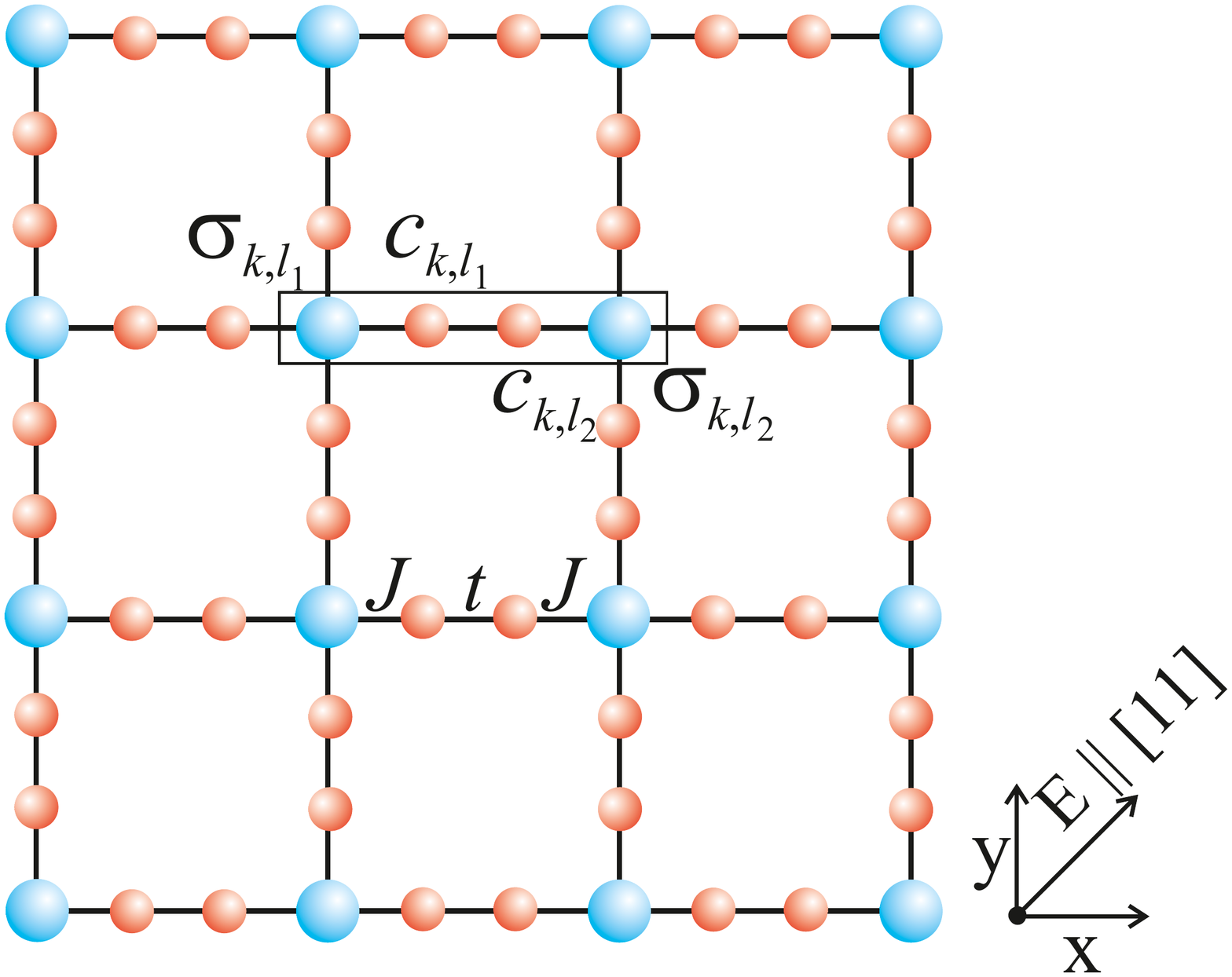}}
 \vspace*{-0.5cm}
\caption{A part of doubly decorated square lattice. Large (blue) balls determine lattice position of the localized Ising spins, while small (red) balls denote decorating sites over which the mobile electrons are delocalized. A rectangle delimits the $k$th bond described by the bond Hamiltonian (\ref{eq1}).}
      \centering
\label{fig0}
\end{figure}

The investigated spin-electron model consists of the localized Ising spins situated at vertices of a square lattice, which are indirectly coupled through the Ising-type exchange interaction mediated by mobile electrons performing a quantum-mechanical hopping at a couple of decorating sites situated on all bonds of a square lattice (see Fig. \ref{fig0}). It is noteworthy that magnetic properties of the coupled spin-electron model on a doubly decorated square lattice have been comprehensively investigated in absence of the external electric field by assuming a quarter filling \cite{tan10}, a half filling \cite{str09,gal11} or a fractional filling \cite{dor14,str15,cen16,cen18} of the atomic orbitals of the decorating sites. The main focus of the present work will be therefore an influence of the external electric field upon magnetic properties of this correlated spin-electron model, which has not been dealt with previously. Because all the considered interactions have only the local character, the total Hamiltonian of the considered spin-electron model can be divided into a sum of the commuting bond Hamiltonians $\hat{\cal H}=\sum_{k=1}^{2N}\hat{\cal H}_k$, whereas the local bond Hamiltonian $\hat{\cal H}_k$ is defined as follows
 \allowdisplaybreaks
\begin{eqnarray}
\hat{\cal H}_k= \!\!\!&-&\!\!\! t(\hat{c}^\dagger_{k,l_1,\uparrow}\hat{c}_{k,l_2,\uparrow}+\hat{c}^\dagger_{k,l_1,\downarrow}\hat{c}_{k,l_2,\downarrow}+
h.c.)\nonumber
\\
\!\!\!&-&\!\!\! J \hat\sigma^z_{k,l_1}(\hat{n}_{k,l_1,\uparrow}-\hat{n}_{k,l_1,\downarrow})-
J\hat\sigma^z_{k,l_2}(\hat{n}_{k,l_2,\uparrow}-\hat{n}_{k,l_2,\downarrow})
\nonumber\\
\!\!\!&-&\!\!\! V^{*}(\hat{n}_{k,l_1,\uparrow}+\hat{n}_{k,l_1,\downarrow}-\hat{n}_{k,l_2,\uparrow}-\hat{n}_{k,l_2,\downarrow})-\mu\hat{n}_{k}
\;.
\label{eq1}
\end{eqnarray}
Above, $\hat{c}^\dagger_{k,l_{\alpha},\gamma}$ and $\hat{c}_{k,l_{\alpha},\gamma}$ ($\alpha$=1,2; $\gamma$=$\uparrow,\downarrow$) denote the creation and annihilation fermionic operators of the mobile electrons from the $k$-th couple of decorating sites, $\hat{n}_{k,l_{\alpha},\gamma}$=$\hat{c}^\dagger_{k,l_{\alpha},\gamma}\hat{c}_{k,l_{\alpha},\gamma}$ and $\hat{n}_k$=$\sum_{\alpha=1,2}\sum_{\gamma=\uparrow,\downarrow}\hat{n}_{k,l_{\alpha},\gamma}$ determine the respective number operators, and $\hat{\sigma}_{k,l_{\alpha}}^z$ represent $z$-component of the Pauli spin matrices corresponding to the localized Ising spins. The first term in Eq.~(\ref{eq1}) determines the kinetic energy of the mobile electrons, while the next two terms stand for the Ising-type exchange interaction between the nearest-neighbor localized Ising spins and mobile electrons. Next, the fourth term takes into account the effect of external electric field $E$ applied along the crystallographic axis \textbf{[11]}. Under this specific spatial orientation, one achieves the same electrostatic potential on horizontal $V_{x}$=$E_{x}|e|d$ and vertical $V_{y}$=$E_{y}|e|d$ bonds ($e$ is an electron charge and $d$ is a distance between decorating atoms), so that one may introduce a new re-scaled electrostatic potential $V^*$=$V/\sqrt{2}$ for the mobile electrons residing on all horizontal and vertical bonds. Finally, the last term involving the chemical potential of the mobile electrons $\mu$ allows to control the electron density on the decorating sites (i.e. the number of mobile electrons per bond).

The grand-canonical partition function of the model under investigation can be exactly calculated by following the procedure elaborated in our previous papers for zero electric field \cite{str15,cen16,cen18}, so it is sufficient to recall just a few most important steps of this procedure and to quote generalized form of the expressions modified by the relevant field term. The grand-canonical partition function can be partially factorized and expressed in terms of the eigenvalues $E_{k,i}$ of the bond Hamiltonian (\ref{eq1})
\begin{eqnarray}
\Xi = \sum_{\{\sigma\}} \prod_{k=1}^{2N} \Xi_k = \sum_{\{\sigma\}} \prod_{k=1}^{2N} \sum_{i=1}^{16} \exp(-\beta E_{k,i}),
\label{eq2}
\end{eqnarray}
where $\beta$ = $1/(k_{\rm B} T)$, $k_{\rm B}$ is the Boltzmann constant, $T$ is the absolute temperature and the summation $\sum_{\{\sigma\}}$ runs over all possible spin states of the localized Ising spins. After tracing out degrees of freedom of the mobile electrons the bond grand partition function $\Xi_k$ solely depends only on spin states of two localized Ising spins, which allows us to replace the bond grand partition function $\Xi_k$ through a simpler expression via the generalized decoration-iteration transformation \cite{fis59,syo72,roj09,str10} 
\begin{eqnarray}
\Xi_k = \sum_{i=1}^{16} \exp(-\beta E_{k,i}) = A\exp(\beta R\sigma_{k,l_1}\sigma_{k,l_2}).
\label{eq3}
\end{eqnarray}
The mapping parameters $A$ and $R$ entering into the decoration-iteration transformation (\ref{eq3}) are 'self-consistently' given by the expressions $A$=$(V_1V_2)^{1/2}$ and $\beta R$=$\frac{1}{2}\ln\left(V_1/V_2\right)$, which are defined through two new functions $V_1$ and $V_2$
\allowdisplaybreaks
\begin{eqnarray}
V_1&=&1+4(z+z^3)\cosh (\beta J) \cosh (\beta t^*)
\nonumber\\&+& 2z^2\left[ 1 + \cosh (2\beta J) + \cosh (2\beta t^*)\right]+z^4\; ,
\nonumber\\
V_2&=&1+2(z+z^3)[\cosh(\beta{B}_+) + \cosh (\beta{B}_-)]
\nonumber\\
&+&2z^2[1+\cosh(\beta {F}_+) + \cosh(\beta {F}_-)] + z^4\;,
\label{eq9}
\end{eqnarray}
where 
\allowdisplaybreaks
\begin{eqnarray}
t^*&=&\sqrt{{V^*}^2+t^2},\;\;\;\;\; {B}_{\pm}=\sqrt{J[J\pm 2V^*]+{t^*}^2},\nonumber\\ 
{F}_{\pm}&=&\sqrt{2\left[J^2+{t^*}^2\right]\pm2
{B}_+{B}_-},\;\;\;\;\;z=e^{\beta\mu}. 
\label{eq9a}
\end{eqnarray}
The electron density, which is defined as the mean value of the overall number operator of the $k$-th bond $\rho\equiv\langle \hat{n}_{k}\rangle$, can be subsequently calculated from the formula
\allowdisplaybreaks
\begin{eqnarray}
\rho\!&=&\!\frac{z}{2N}\frac{\partial \ln \Xi}{\partial z}\!=\!
\frac{z}{2}\left\{\frac{V'_{1}}{V_1}(1+\varepsilon)+\frac{V'_{2}}{V_2}(1-\varepsilon)\right\}\!,
\label{eq10}
\end{eqnarray}
where $\varepsilon \equiv\langle\sigma_{k,l_1}\sigma_{k,l_2}\rangle$ denotes the pair correlation function between the nearest-neighbor localized Ising spins and $V'_{1}$=$\partial V_1/\partial z$, $V'_{2}$=$\partial V_2/\partial z$. The last equation plays the crucial role in the computational process, because it may be viewed as the equation of state when defining correspondence between two conjugated variables - the chemical potential and electron density of the mobile electrons. 

The electric polarization can be related to an elementary electric dipole moment $P$ of the mobile electrons, which can be expressed as a difference of the mean electron concentrations at two decorating sites $P=\sum_{\gamma=\uparrow,\downarrow} \langle \hat{n}_{k,l_{1},\gamma}\rangle-\langle\hat{n}_{k,l_{2},\gamma}\rangle$. An existence of the spontaneous ferromagnetic and antiferromagnetic order can be inferred from previous studies of the zero-field case \cite{dor14,str15,cen16,cen18} and hence, it might be useful to calculate the uniform magnetization of the localized Ising spins $m_i=(\langle \hat{\sigma}_{k,l_1}^z \rangle+\langle\hat{\sigma}_{k,l_2}^z\rangle)/2$ and the mobile electrons $m_e$ per bond as the order parameters for the ferromagnetic phase along with the staggered magnetization of the localized Ising spins $m^s_i=(\langle \hat{\sigma}_{k,l_1}^z \rangle-\langle\hat{\sigma}_{k,l_2}^z\rangle)/2$  and the mobile electrons $m^s_e$ as the order parameters of the antiferromagnetic phase. Exact mapping theorems \cite{bar88,bar91} would imply that the uniform and staggered magnetizations of the localized Ising spins directly equal to the single-site uniform and staggered magnetizations of the effective Ising model $m_i = m_{IM} (\beta, R>0)$ and $m_i^s = m_{IM}^s (\beta, R<0)$, while the uniform and staggered magnetizations of the mobile electrons follow from the relations
\allowdisplaybreaks
\begin{eqnarray}
m_{e}\!\!\!&=&\!\!\!\Bigg\langle \sum_{j=1,2}\frac{\partial \ln \Xi_k}{\partial \beta J\sigma_{k,l_j}}\Bigg\rangle=m_i\frac{W_1}{V_1},\\
{m}^s_{e}\!\!\!&=&\!\!\!\Bigg\langle\!\sum_{j=1,2}\!\!\!\frac{(-1)^{j+1}\partial \ln \Xi_k}{\partial \beta J\sigma_{k,l_{j}}}\!\Bigg\rangle
\!=\!\frac{T_1}{V_1}\left(m_i\!+\!m_i^s\right)\!+\!\frac{T_2}{V_2}m_i^s\!,
\label{eq13}
\end{eqnarray}
where we have used the abbreviations $W_1$, $T_1$ and $T_2$ for the following functions
\allowdisplaybreaks
\begin{eqnarray}
{W_1}&=&4(z+z^3)\cosh(\beta t^*)\sinh(\beta J) +4z^2\sinh (2\beta J),\nonumber\\
{T_1}&=&4(V^*/t^*)(z-z^3)\sinh(\beta J)\sinh(\beta t^*),
 \label{eq14}\\
{T_2}&=& 2(z+z^3)[(J+V^*)SB_+ + (J-V^*)SB_-]
\nonumber\\
\!\!\!&\hspace{-1cm}+&\!\!\!\hspace{-0.5cm}4Jz^2[(1\!\!+\!\!\frac{J^2\!-\!{V^*}^2\!+\!t^2}{{B}_+{B}_-})SF_+ \!\!+\!\!(1\!\!-\!\!\frac{J^2\!-\!{V^*}^2\!+\!t^2}{{B}_+{B}_-})SF_-].
\nonumber
\end{eqnarray}
To save the expression lucidity, the novel notations $SB_{\pm}$=$(\sinh \beta B_{\pm})/B_{\pm}$ and  $SF_{\pm}$=$(\sinh \beta F_{\pm})/F_{\pm}$ are defined.

\section{Results and discussion}
\label{sec:result}

\begin{figure}[h!]%
{\includegraphics[width=.35\textwidth,height=4.8cm,trim=2.5cm 9cm 4.5cm 7.5cm, clip]{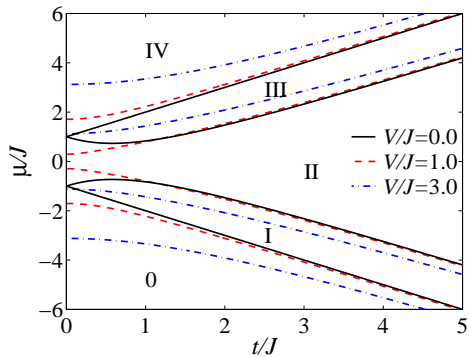}}
\caption{The ground-state phase diagram in the $t/J-\mu/J$ plane as constructed from the lowest-energy eigenstates of the bond Hamiltonian (\ref{eq1}) for  three different values of the external electric field $V/J$.}
     \vspace*{0.5cm}
     \centering
\label{fig1}
\end{figure}

Let us start our discussion of the most interesting results with a comprehensive analysis of the ground-state phase diagram, which was determined from the lowest-energy eigenstates of the bond Hamiltonian (\ref{eq1}). Five different ground states either with a paramagnetic (P), ferromagnetic (F) or antiferromagnetic (AF) spin arrangements can be detected in the overall parameter space depending on a competition between the hopping term $t>0$, the ferromagnetic exchange constant $J>0$, the electric field $V>0$ and the chemical potential $\mu$. It should be pointed out that the applied electric field relevantly influences only the electron subsystem and consequently, the respective probability amplitudes of individual  basis states of the mobile electrons are modified by the electric field in contrast with the basis states of the localized Ising spins that are unaffected and remain completely identical as in the zero-field case $V$=$0$ \cite{str15}. In absence of the external electric field a homogeneous distribution of the mobile electrons over the pairs of decorating sites is realized in the ferromagnetic phases I and III with odd number of the mobile electrons per bond \cite{str15}, while a charge segregation at one of two decorating sites takes place in the ferromagnetic phases I and III due to the nonzero electric field as exemplified by the eigenvectors
\begin{eqnarray}
|\mbox{I}\rangle \!\!\!&=&\!\!\! \prod_{k=1}^{2N} |1\rangle_{\sigma_{k,l_1}}\otimes \left(\cos\alpha
|\!\uparrow,0\rangle_k+\sin\alpha|0,\uparrow\rangle_k\right)\otimes|1\rangle_{\sigma_{k,l_2}}, \nonumber \\  
|\mbox{III}\rangle \!\!\!&=&\!\!\! \prod_{k=1}^{2N} |1\rangle_{\sigma_{k,l_1}}\!\!\!\otimes\! \left(\cos\alpha|\!\uparrow\downarrow,\uparrow\rangle_k-\sin\alpha|\!\uparrow,\uparrow\downarrow\rangle_k\right)\!\otimes\!|1\rangle_{\sigma_{k,l_2}}, \nonumber \\
\label{eq:gsf}
\end{eqnarray}
where the emergent quantum superposition of two basis states of the mobile electrons is determined through the mixing angle $\alpha$ defined by $\tan\alpha$=$(t^*\!\!-\!\!V^*)/t$. The markedly different situation appears in the quantum antiferromagnetic phase II emergent around the half-filled band case, where the classical N\'eel order of the localized Ising spins is accompanied with a quantum superposition of two magnetic and two non-magnetic ionic states of the mobile electrons \cite{str15}. The magnetism of the electron subsystem basically changes under the influence of the external electric field, which significantly contributes to a quantum reduction of the spontaneous staggered magnetization of the mobile electrons according to the eigenvector 
\begin{eqnarray}
|\mbox{II}\rangle=\prod_{k=1}^{2N}|1\rangle_{\sigma_{k,l_1}}\!\otimes\!\left[ \right. \!\!\!\!\!\!&&\!\!\!\!\!\! 
\alpha_2|\!\uparrow,\downarrow\rangle_k\! + \beta_2|\!\downarrow,\uparrow\rangle_k \!+\!\gamma_2|\!\uparrow\downarrow,0\rangle_k\! 
\nonumber \\ 
\!\!\!&+&\!\!\! \left. \delta_2|0,\uparrow\downarrow\rangle_k\right]\!\otimes\!|\!-1\rangle_{\sigma_{k,l_2}}, 
\label{eq:gsaf}
\end{eqnarray}
which involves a quantum superposition of four aforementioned electronic states defined through the mixing angles $\alpha_2$=$2t\eta_3(\eta_3\!-\!\eta_1)/\eta_4$, $\beta_2$=$\frac{-\alpha_2(\eta_1+\eta_3)}{\eta_3-\eta_1}$, $\gamma_2$=$\frac{\alpha_2(\eta_1+\eta_3)(\eta_3-\eta_2)}{2t\eta_3}$ and $\delta_2$=$\gamma_2\frac{\eta_2+\eta_3}{\eta_3-\eta_2}$, $\eta_1$=-$2J$, $\eta_2$=-$2V^*$, $\eta_3$=${F}_+$,$\eta_4$=$\sqrt{2}\sqrt{4t^2\eta_3^2(\eta_3^2+\eta_1^2)+(\eta_3^2+\eta_2^2)(\eta_3^2-\eta_1^2)^2}$. It is quite clear that the magnetism of two paramagnetic phases 0 and IV with empty or fully occupied decorating sites is not influenced by the external electric field and hence, the eigenvectors remain completely the same as predicted in our earlier work \cite{str15}. 
 Another important observation is that the increasing $V$ gives to rise an effective antiferromagnetic coupling between the localized spins instead of the ferromagnetic one, whereas for sufficiently large $V/J$ a clear enhancement of the parameter region corresponding to the phase II is observed (see Fig.~\ref{fig1}).

\begin{figure}[htb]
{\includegraphics[width=0.5\textwidth]{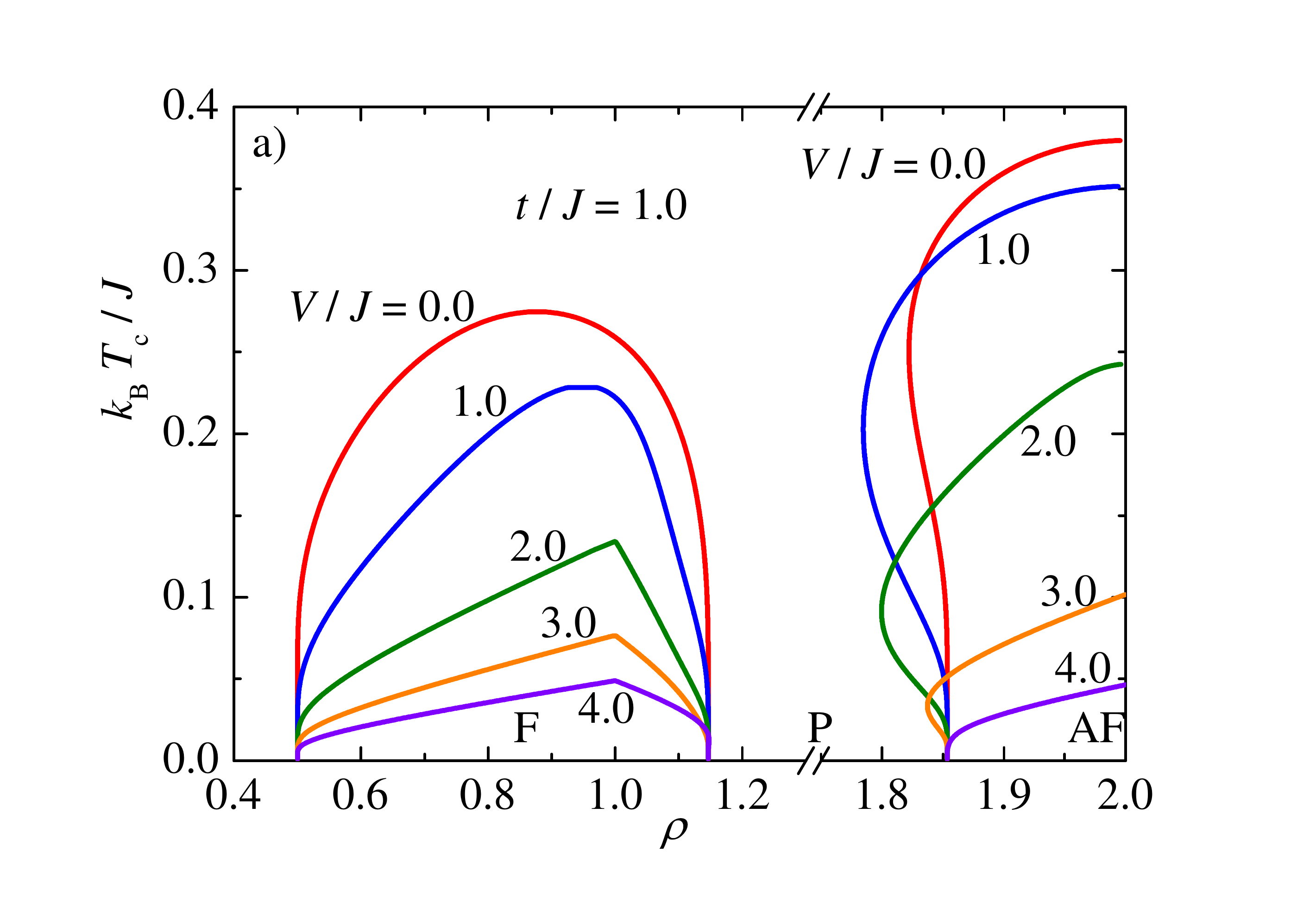}}\\
\includegraphics[width=0.5\textwidth]{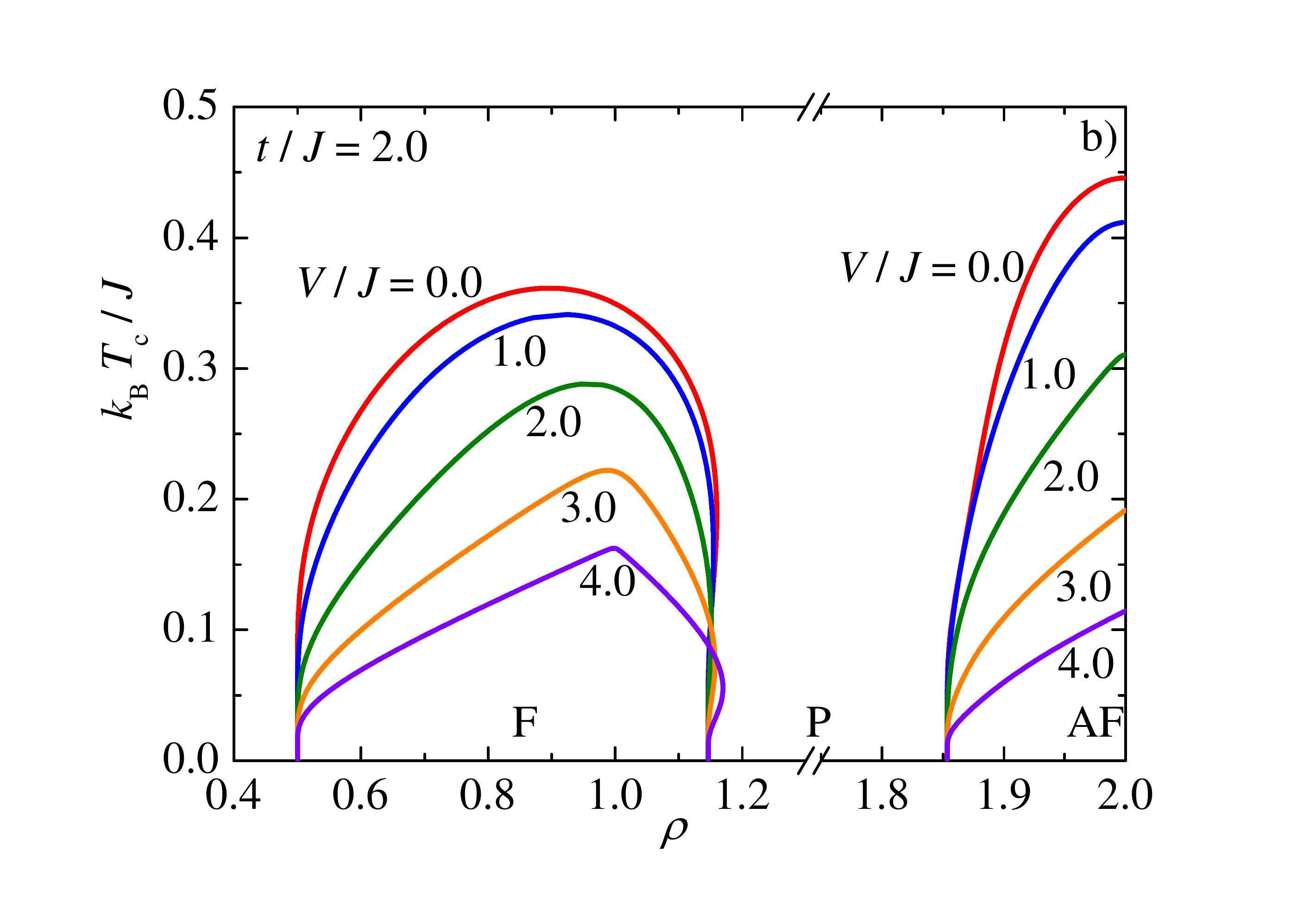}
\caption{Finite-temperature phase diagrams in the form of critical temperature versus electron density plot obtained for several values of the external electric field $V/J$ and two different values of the hopping term: (a) $t/J = 1.0$; (b) $t/J = 2.0$. Note that the dome of critical temperatures emergent at electron densities close to a quarter (half) filling corresponds to the spontaneously ordered ferromagnetic (antiferromagnetic) phase, while the disordered paramagnetic phase extends over moderate values of the electron densities predominantly skipped through the axis break.}
\vspace*{-0.5cm}
\centering
\label{fig2}
\end{figure}

Next, let us investigate a stability of the spontaneous ferromagnetic and antiferromagnetic magnetic order with respect to temperature and the external electric field upon varying the electron density $\rho\in(0,2)$ at the decorating sites (the critical temperature in the other range $\rho\in(2,4)$ is symmetrical due to a particle-hole symmetry). It has been demonstrated that the spontaneous ferromagnetic and antiferromagnetic order is realized at a quarter filling \cite{tan10} and a half filling \cite{str09}, respectively, whereas the spontaneous ferromagnetic and antiferromagnetic orders are still preserved when the electron density slightly deviates from the quarter- and half-filling band case \cite{dor14,str15}. It is quite evident from the finite-temperature phase diagrams shown in Fig.~\ref{fig2} that the applied electric field generally reduces a critical temperature of the ferromagnetic and antiferromagnetic phase, while the electron concentrations at which the spontaneous ferromagnetic and antiferromagnetic orders emerge remain unaffected. The reduction of critical temperature of the ferromagnetic phase owing to the external electric field can be related to a greater localization of the mobile electron at one of two decorating sites, which consequently transmits the effective ferromagnetic coupling between the localized Ising spins less effectively. Similarly, the critical temperature of the antiferromagnetic phase also shrinks upon increasing of the external electric field, because the electric field supports a charge segregation of the mobile electrons at one of the decorating sites and thus reduces the effective antiferromagnetic interaction transmitted by two mobile electrons residing the same bond. 

Another interesting observation, which directly follows from the finite-temperature phase diagram shown in Fig.~\ref{fig2}, is an existence of a sharp kink in the dome of critical temperature delimiting a stability region of the ferromagnetic phase. The sharp kink can be observed close to a quarter filling on assumption that the electric field is sufficiently strong with respect to the hopping term $V \gtrsim 2 t$. This effect can be attributed to the competitive influence of the kinetic term and electric field: the hopping term $t$ favours a homogeneous distribution of the mobile electrons over the pairs of decorating sites, while the electric field contrarily acts in favour of the charge separation. Moreover, it can be seen from Fig.~\ref{fig2} that the electric field gives rise to reentrant phase transitions either of the antiferromagnetic phase (Fig.~\ref{fig2}a) or the ferromagnetic phase (Fig.~\ref{fig2}b). While the reentrant phase transitions of the antiferromagnetic phase emergent below $\rho \lesssim 1.854$ can be already found in absence of the external electric field \cite{str15} and the relevant field term may just spread reentrance over a wider interval of the electron densities, the reentrant phase transitions observable slightly above the upper edge $\rho \gtrsim 1.146$ of the ferromagnetic dome arise from the external electric field as they are totally absent in the zero-field case (see Fig.~\ref{fig2}b). 

\begin{figure}[h]%
{\includegraphics*[width=.45\textwidth,height=5.5cm,trim=2.5cm 9.cm 3.5cm 8cm, clip]{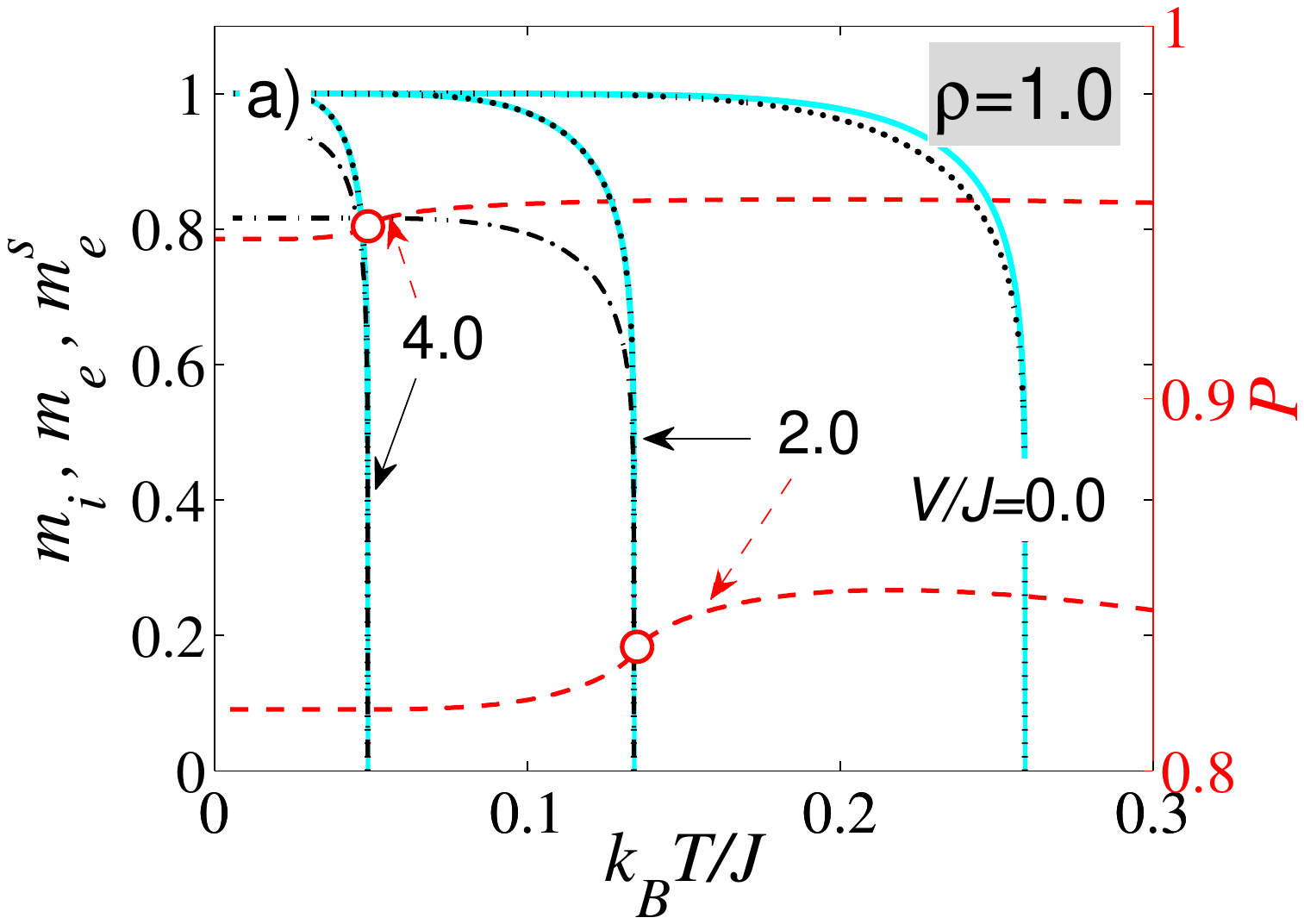}}
{\includegraphics*[width=.45\textwidth,height=5.5cm,trim=2.5cm 9cm 3.5cm 8cm, clip]{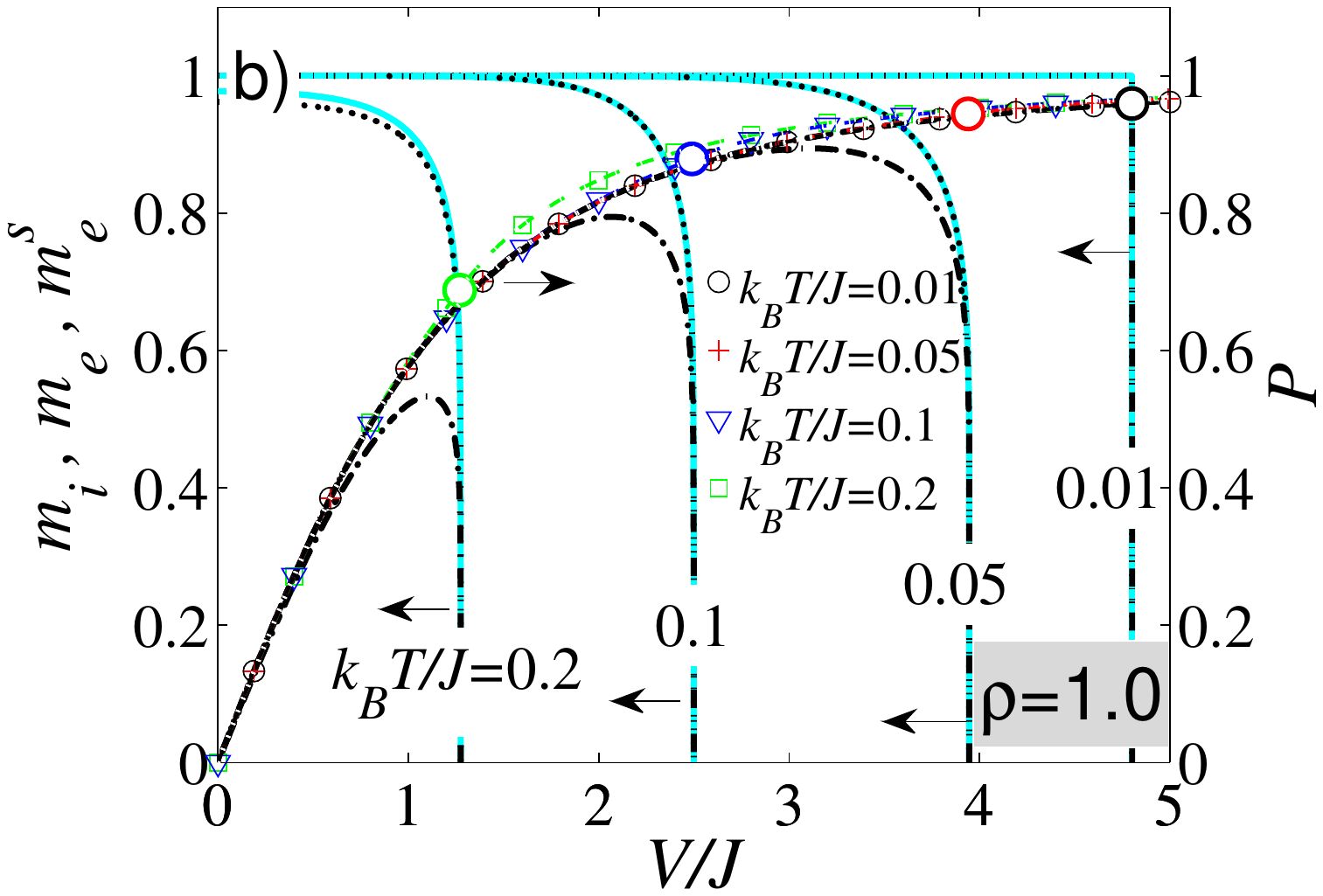}}
\caption{ (a) Thermal and (b) electric-field variations of the spontaneous uniform magnetization $m_i$ (solid lines), $m_e$ (dotted lines), the staggered magnetization of the mobile electrons $m_e^s$ (dashed-dotted lines) and the electric polarization $P$ (dashed lines with/without symbols) at a quarter filling $\rho$=$1$ and $t/J=1$ for selected model parameters. Open circles denote weak-singular points of the electric polarization.}
     \vspace*{0.5cm}
     \centering
\label{fig3}
\end{figure}

\begin{figure}[h]%
{\includegraphics*[width=.45\textwidth,height=5.5cm,trim=2.5cm 9.cm 3.5cm 8cm, clip]{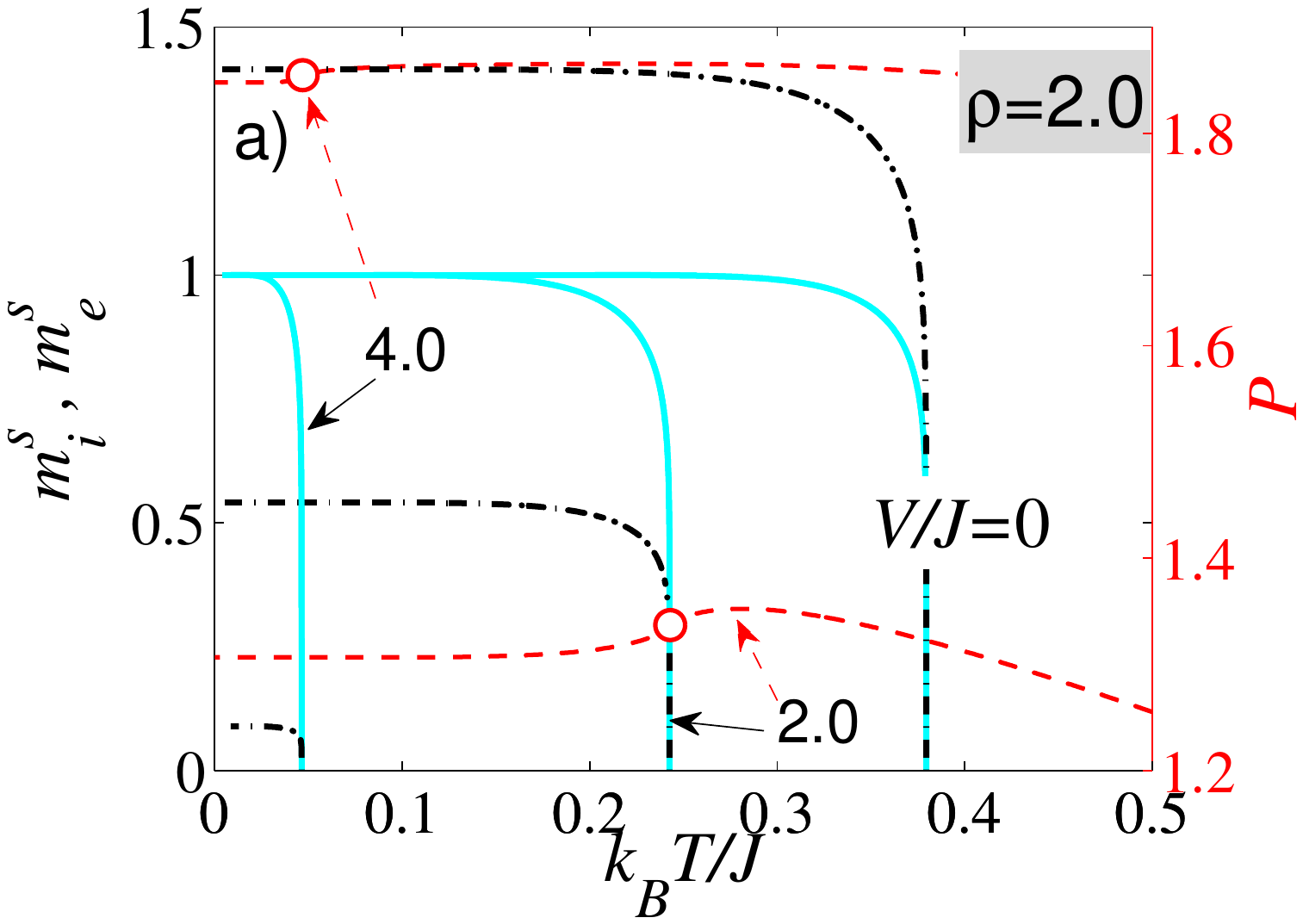}}
{\includegraphics*[width=.45\textwidth,height=5.5cm,trim=2.5cm 9cm 3.5cm 8cm, clip]{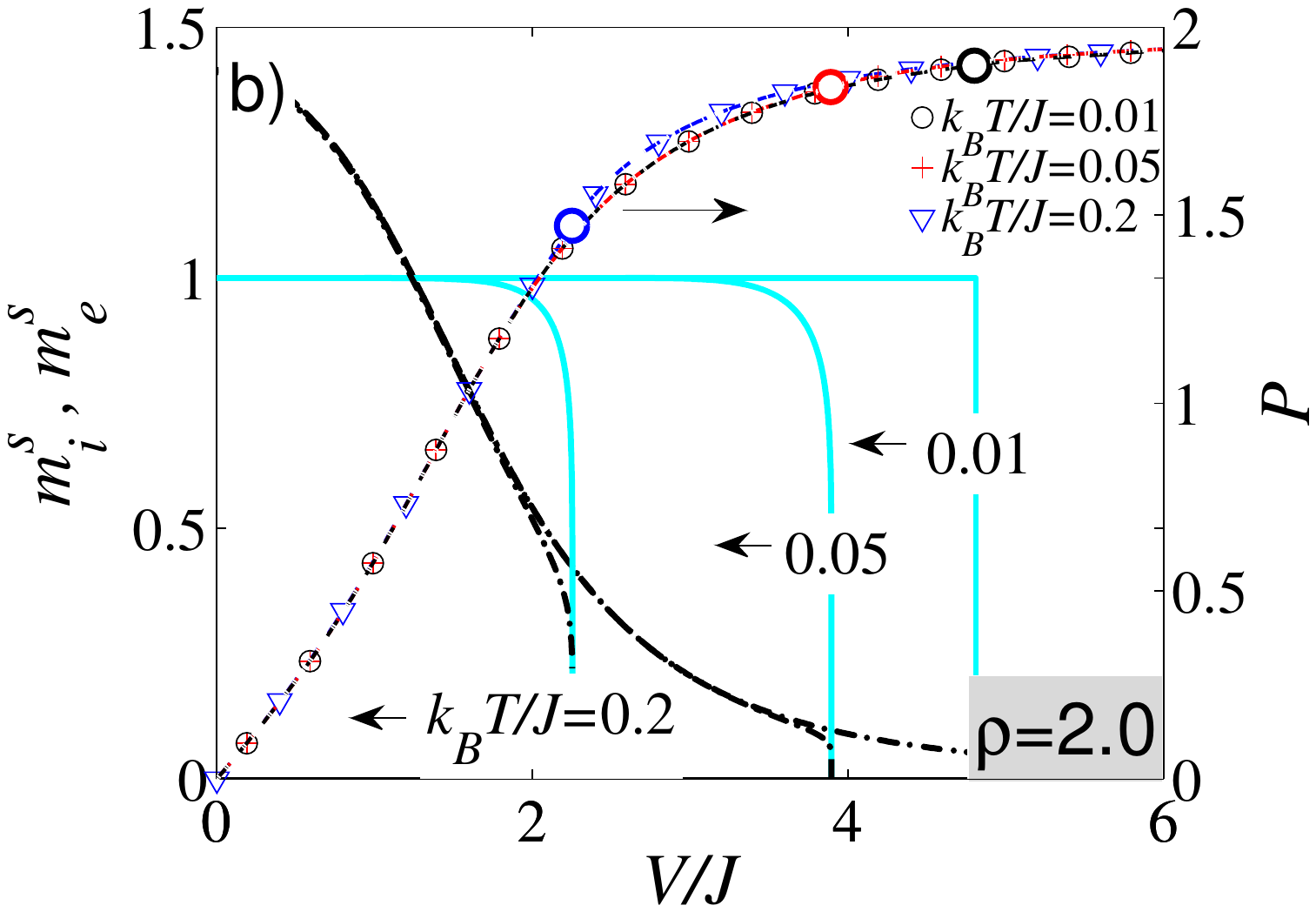}}
\caption{(a) Thermal and (b) electric-field variations of the spontaneous staggered magnetizations $m_i^s$ (solid lines), $m^s_e$ (dashed-dotted lines) and the electric polarization $P$ (dashed lines with/without symbols) at a half filling $\rho$=$2$ and $t/J=1$ for selected model parameters. Open circles denote weak-singular points of the electric polarization.}
     \vspace*{0.5cm}
     \centering
\label{fig4}
\end{figure}

Last but not least, let us analyze in detail the spontaneous magnetization and electric polarization under a concurrent influence of temperature and electric field at two different electron densities. For illustration, we have chosen two electron concentrations equal to a quarter- and half-filling band case, which demonstrate typical thermal behavior of the ferromagnetic and antiferromagnetic phase, respectively. It is quite evident from Figs.~\ref{fig3},\ref{fig4} that there exists a strong correlation between the electric and magnetic properties of the investigated system. First, let us make a few comments on the typical behavior of the spontaneous magnetization and electric polarization of the ferromagnetic phase. It can be seen from Fig.~\ref{fig3}a) that the sublattice magnetizations of the localized Ising spins and mobile electrons exhibit standard temperature dependencies with a steep power-law decline observable in a vicinity of the critical temperature, while the electric polarization displays a more striking temperature dependence with a weak singularity located at the critical temperature. It should be also mentioned that the electric field gradually increases the staggered magnetization of the mobile electrons despite the spontaneous ferromagnetic order. Moreover, it is quite obvious from Fig.~\ref{fig3}b) that the electric field  gradually destroys the spontaneous magnetizations of the localized Ising spins and mobile electrons, whereas the relevant electric-field dependence just smears out upon increasing of temperature. It could be thus concluded that an enhanced magnetoelectric effect can be detected in vicinity of a critical point related to a continuous phase transition of the ferromagnetic phase. On the other hand, the electric polarization displays a monotonous increase with the external electric field even though it is still displays a weak singular point at the relevant critical point. It is quite remarkable that the staggered magnetization of the mobile electrons closely follows the electric polarization at low enough electric fields, but then it tend to vanish together with the spontaneous uniform magnetizations of the localized Ising spins and mobile electrons.   

As far as the thermal behavior of the antiferromagnetic phase is concerned, the staggered magnetizations of the localized Ising spins and mobile electrons still display standard thermal dependencies with a steep power-law decline emerging in a vicinity of the critical temperature (see Fig.~\ref{fig4}a)). It is worthwhile to remark, however, that the zero-temperature asymptotic value of the staggered magnetization of the mobile electrons is gradually suppressed upon increasing of the electric field due to a charge segregation. Similarly to the previous case, the electric polarization exhibits a nonmonotonous thermal dependence with a weak singularity at the critical temperature. It is quite apparent from Fig.~\ref{fig4}b) that the spontaneous staggered magnetizations of the localized Ising spins and mobile electrons disappear upon increasing of the electric field, whereas the relevant field dependence gradually smears out upon increasing temperature. Thus, one may observe an enhanced magnetoelectric effect related to a vigorous change of the spontaneous staggered magnetization driven by the external electric field close to a continuous phase transition of the antiferromagnetic phase. The electric polarization contrarily exhibits a smooth monotonous electric-field dependence with a weak singular point located at the critical temperature. 
\section{Summary and conclusions}
\label{sec:conc}

In the present work we have rigorously studied the ground-state and finite-temperature phase diagrams of a coupled spin-electron model on a doubly decorated square lattice in an external electric field applied along the crystallographic axis \textbf{[11]} together with the temperature and electric-field dependencies of the spontaneous magnetizations and electric polarization. It has been shown that the ground-state phase diagram still involves five different ground states either with ferromagnetic, antiferromagnetic or paramagnetic spin arrangements, which have already been reported for the zero electric field \cite{str15}. Although the applied electric field does not produce any new ground state it has a nonnegligible effect upon the stability, magnetic and electric features of the relevant ground states through the charge separation inducing the electric polarization. It has been verified that the external electric field favours at zero temperature the antiferromagnetic phase before the ferromagnetic one. However, the external electric field generally suppress the spontaneous ferromagnetic and antiferromagnetic orders at finite temperature as evidenced by the reduction of critical temperatures of both these phases. While the spontaneous magnetizations of the localized Ising spins and mobile electrons display standard temperature dependencies with a steep power-law dependence at the critical temperature, the electric polarization exhibits a non-monotonous temperature dependence with a remarkable weak singularity located at the critical temperature. However, the most interesting finding of the present work concerns with an observation of an enhanced magnetoelectric effect in vicinity of a critical temperature of a continuous phase transition of the ferromagnetic or antiferromagnetic phase, at which a substantial reduction of the spontaneous uniform or staggered magnetization can be achieved by the external electric field. 
 
\begin{acknowledgments}
This work was financially supported by the grant of the Slovak Research and Development Agency provided under the contract No. APVV-16-0186 and by the grant of The Ministry of Education, Science, Research, and Sport of the Slovak Republic provided under the contract No. VEGA 1/0043/16. H.C. also acknowledges support by the ERDF EU Grant under contract No. ITMS  26220120047.
\end{acknowledgments}

\end{document}